\documentclass[12pt]{article}
\begin{document}
\baselineskip=16pt
\begin{titlepage}
\setcounter{page}{0}
\begin{center}

\vspace{0.5cm} {\Large \bf Cosmological Scaling Solutions and
Multiple Exponential Potentials }\\
\vspace{10mm}
Zong-Kuan Guo\footnote{e-mail address: guozk@itp.ac.cn}$^{a}$,
Yun-Song Piao\footnote{e-mail address: yspiao@itp.ac.cn}$^{a}$,
and
Yuan-Zhong Zhang$^{b,a}$ \\
\vspace{6mm} {\footnotesize{\it
  $^a$Institute of Theoretical Physics, Chinese Academy of Sciences,
      P.O. Box 2735, Beijing 100080, China\\
  $^b$CCAST (World Lab.), P.O. Box 8730, Beijing 100080\\}}

\vspace*{5mm} \normalsize
\smallskip
\medskip
\smallskip
\end{center}
\vskip0.6in \centerline{\large\bf Abstract}
{We present a phase-space analysis of cosmology containing multiple scalar
fields with positive and negative exponential potentials. We show that
there exist power-law multi-kinetic-potential scaling solutions for
sufficiently flat positive potentials or steep negative potentials. The
former is the unique late-time attractor and the well-known assisted
inflationary solution, but the later is never unstable in an expanding
universe. Moreover, for steep negative potentials there exist a
kinetic-dominated regime in which each solution is a late-time attractor.
We briefly discuss the physical consequences of these
results.}

\vspace*{2mm}
\end{titlepage}

Scalar field cosmological models are of great importance in modern
cosmology. The dark energy is attributed to the dynamics of a
scalar field, which convincingly realizes the goal of explaining
present-day cosmic acceleration generically using only attractor
solutions~\cite{CD}. A scalar field can drive an accelerated
expansion and thus provides possible models for cosmological
inflation in the early universe~\cite{GU}. In particular, there
have been a number of studies of spatially homogeneous scalar
field cosmological models with an exponential potential. There are
already known to have interesting properties; for example, if one
has a universe containing a perfect fluid and such a scalar field,
then for a wide range of parameters the scalar field mimics the
perfect fluid, adopting its equation of state~\cite{CL}. These
scaling solutions are attractors at late times~\cite{ST}. The
inflation~\cite{LM, PCZ} and other cosmological effect~\cite{HL}
of multiple scalar fields have also been considered.

The scale-invariant form makes the exponential potential
particularly simple to study analytically. There are well-known
exact solutions corresponding to power-law solutions for the
cosmological scale factor $a\propto t^p$ in a spatially flat
Friedmann-Robertson-Walker (FRW) model~\cite{FLM},
but more generally the coupled
Einstein-Klein-Gordon equations for a single field can be reduced
to a one-dimensional system which makes it particularly suited to
a qualitative analysis~\cite{JJH,CLW}. Recently, adopting a system
of dimensionless dynamical variables~\cite{EW},
the cosmological scaling solutions with positive and negative
exponentials has been studied~\cite{HW}. Usually there are many
scalar fields with exponential potentials in supergravity,
superstring and the generalized Einstein theories, thus multi
potentials may be more important.
In this paper, we will consider multiple scalar fields with
positive and negative exponential potentials. We have assumed that
there is no direct coupling between the exponential potentials.
The only interaction is gravitational. A phase-space analysis of
the spatially flat FRW models shows that there exist cosmological
scaling solutions which are the unique late-time attractors, and
successful inflationary solutions which are driven by multiple
scalar fields with a wide range of each potential slope parameter
$\lambda$.

We start with more general model with $m$ scalar
fields $\phi _i$, in which each has an identical-slope potential
\begin{equation}
V_i(\phi _i)=V_{0i}\exp{(-\lambda \kappa \phi _i)}
\end{equation}
where $\kappa ^2\equiv 8 \pi G_N$ is the gravitational coupling
and $\lambda$ is a dimensionless constant characterising the slope
of the potential.
Note that there is no direct coupling of the fields, which
influence each other only via their effect on the expansion.
The evolution equation of each scalar field for a
spatially flat FRW model with Hubble parameter $H$ is
\begin{equation}
\label{2E}
\ddot{\phi _i}+3H\dot{\phi _i}+\frac{dV_i(\phi _i)}{d\phi _i}=0
\end{equation}
subject to the Friedmann constraint
\begin{equation}
\label{3E}
H^2=\frac{\kappa ^2}{3}\sum_{i=1}^{m}\left[V_i(\phi
_i)+\frac{1}{2}\dot{\phi}_i^2\right]
\end{equation}
Defining $2m$ dimensionless variables
\begin{equation}
x_i=\frac{\kappa \dot{\phi _i}}{\sqrt{6}H}, \qquad
y_i=\frac{\kappa \sqrt{|V_i|}}{\sqrt{3}H}
\end{equation}
the evolution equations (\ref{2E}) can be written
as an autonomous system:
\begin{eqnarray} 
\label{5E} 
x'_i&=&-3x_i\left(1-\sum_{j=1}^{m}x_j^2\right)
\pm \lambda \sqrt{\frac{3}{2}}y_i^2 \\ 
y'_i&=&y_i\left(3\sum_{j=1}^{m}x_j^2-\lambda \sqrt{\frac{3}{2}}x_i\right) 
\label{6E} 
\end{eqnarray} 
where a prime denotes a derivative with respect to the logarithm of
the scalar factor, $N\equiv \ln a$, and the constraint equation
(\ref{3E}) becomes
\begin{equation}
\label{7CE}
\sum_{i=1}^{m}\left(x_i^2\pm y_i^2\right)=1
\end{equation}
Throughout we will use upper/lower signs to
denote the two distinct cases of $\pm V_i>0$.
$x_i^2$ measures the contribution to the expansion due to the field's
kinetic energy density, while $\pm y_i^2$ represents the contribution
of the potential energy. We will restrict our discussion of the existence
and stability of critical points to expanding universes with $H>0$, i.e.,
$y \ge 0$, and $\lambda > 0$.
Critical points correspond to fixed points where $x'_i=0$,
$y'_i=0$, and there are self-similar solutions with
\begin{equation}
\frac{\dot{H}}{H^2}=-3\sum_{i=1}^{m}x_i^2
\end{equation}
This corresponds to an expanding universe with a scale factor $a(t)$
given by $a\propto t^p$ or a contracting universe with a scalar
factor
given by $a\propto (-t)^p$, where 
\begin{equation}
p=\frac{1}{3\sum_{i=1}^{m}x_i^2}
\label{SFP}
\end{equation}
The system (\ref{5E}) and (\ref{6E}) has at most one $m$-dimensional
sphere embedded in $2m$-dimensional phase-space corresponding to
kinetic-dominated solutions, and $(2^m-1)$ fixed points, one of
which is a $m$-kinetic-potential scaling solution.

In order to analysis the stability of the critical points, we only
consider the cosmologies containing two scalar fields. There are
one unit circle $S$ and three fixed points $A_1$, $A_2$ and $B$ listed
in Table 1.
Using the Friedmann constraint equation (\ref{7CE}), we reduce
Eqs.(\ref{5E}) and (\ref{6E}) to three independent equations
\begin{eqnarray} \label{IE1}
x'_1 &=& -3x_1 (1-x_1^2-x_2^2) \pm \lambda \sqrt{\frac{3}{2}}y_1^2 \\ 
x'_2 &=& -3x_2 (1-x_1^2-x_2^2)+\lambda \sqrt{\frac{3}{2}}
(1-x_1^2-x_2^2 \mp y_1^2) \\ 
y'_1 &=& y_1(3x_1^2+3x_2^2-\lambda \sqrt{\frac{3}{2}}x_1) \label{IE3}
\end{eqnarray}
Substituting linear perturbations about the critical
points $x_1 \to x_1+\delta x_1$, $x_2 \to x_2+\delta x_2$ and
$y_1 \to y_1+\delta y_1$ into Eqs.(\ref{IE1})-(\ref{IE3}),
to first-order in the perturbations, gives equations of motion
\begin{equation} \label{EM}
\left(
\begin{array}{c}
\delta x_1' \\
\delta x_2' \\
\delta y_1'
\end{array}
\right) = {\cal M} \left(
\begin{array}{c}
\delta x_1 \\
\delta x_2 \\
\delta y_1
\end{array}
\right) 
\end{equation}		
where
\begin{equation}
{\cal M}=\left(
\begin{array}{ccc}
-3+9x_1^2+3x_2^2 & 6x_1 x_2 & \pm \lambda \sqrt{6}y_1 \\
6x_1 x_2-\lambda \sqrt{6} x_1 & 
-3+3x_1^2+9x_2^2-\lambda \sqrt{6}x_2 & \mp \lambda \sqrt{6}y_1 \\
6x_1 y_1-\lambda \sqrt{\frac{3}{2}}y_1 & 6 x_2 y_1 & 
3x_1^2+3 x_2^2-\lambda \sqrt{\frac{3}{2}}x_1 
\end{array}
\right) 
\end{equation}
The general solution for the evolution of linear perturbations can be
written as
\begin{eqnarray} 
\delta x_1 &=& u_1\exp(m_1 N)+u_2\exp(m_2 N)+u_3\exp(m_3 N) \nonumber \\
\delta x_2 &=& v_1\exp(m_1 N)+v_2\exp(m_2 N)+v_3\exp(m_3 N) \label{EV} \\
\delta y_1 &=& w_1\exp(m_1 N)+w_2\exp(m_2 N)+w_3\exp(m_3 N) \nonumber
\end{eqnarray}
where $m_1$, $m_2$ and $m_3$ are the eigenvalues of the matrix ${\cal M}$.
Thus stability requires the real part of all eigenvalues being
negative.

\textbf{S:} $x_1^2+x_2^2=1$, $y_1=y_2=0$ \\
These kinetic-dominated solutions exist for any form of the
potential, which are equivalent to stiff-fluid dominated evolution
with $a\propto t^{1/3}$ irrespective of the nature of the
potential, with the eigenvalues
\begin{eqnarray*}
m_1 &=& 0 \\
m_2 &=& 3-\sqrt{\frac{3}{2}} \lambda x_1 \\
m_3 &=& 6-\sqrt{6} \lambda x_2
\end{eqnarray*}
Thus the solutions are stable to potential energy perturbations
for $\lambda x_1 > \sqrt{6}$ and $\lambda x_2
> \sqrt{6}$. Using the constraint equation
(\ref{7CE}) we find $1 \ge 2x_1 x_2>12/ \lambda^2$. That is,
there exist stable points only for sufficiently steep
$(\lambda >2\sqrt{3})$ potential.

\textbf{A$_1$:} $x_1=\frac{\lambda}{\sqrt{6}}$, $y_1=\sqrt{\pm
(1-\frac{\lambda
^2}{6})}$, $x_2=y_2=0$

\textbf{A$_2$:} $x_1=y_1=0$, $x_2=\frac{\lambda}{\sqrt{6}}$, $y_2=\sqrt{\pm
(1-\frac{\lambda
 ^2}{6})})$ \\
The two single-potential-kinetic solutions exist for sufficiently
flat $(\lambda ^2<6)$ positive potentials or steep $(\lambda
^2>6)$ negative potentials. The power-law exponent, $p=2/\lambda
^2$, depends on the slope of the potential. From Eq.(\ref{EM}) we find
the eigenvalues
\begin{eqnarray*}
m_1 &=& \lambda ^2 \\ 
m_2 &=& \frac{1}{2} (\lambda ^2-6) \\
m_3 &=& \frac{1}{2} (\lambda ^2-6)
\end{eqnarray*}
Thus the single-potential-kinetic solutions are unstable for the
positive and negative potentials. This indicates that the stability
is destroyed by the potential energy perturbations of another
scalar field.

\textbf{B:} $x_1=x_2=\frac{\lambda}{2\sqrt{6}}$,
$y_1=y_2=\sqrt{\pm
(\frac{1}{2}-\frac{\lambda ^2}{24})}$ \\
The double-potential-kinetic scaling solution exist for flat
$(\lambda ^2<12)$ positive potentials, or steep $(\lambda ^2>12)$
negative potentials. This corresponds to a power-law solution with
$a\propto t^{4/\lambda ^2}$. Linear perturbations yield three
eigenvalues
\begin{eqnarray*}
m_1 &=& \frac{1}{4}(\lambda ^2-12) \\
m_2 &=& \frac{1}{8}(\lambda ^2-12)-
        \frac{3}{8}\sqrt{(\lambda ^2-12)(\lambda ^2-4/3)} \\
m_3 &=& \frac{1}{8}(\lambda ^2-12)+
        \frac{3}{8}\sqrt{(\lambda ^2-12)(\lambda ^2-4/3)}
\end{eqnarray*}
For $4/3<\lambda ^2<12$, $\sqrt{(\lambda ^2-12)(\lambda ^2-4/3)}$ is
replaced by $i\sqrt{(\lambda ^2-12)(4/3-\lambda ^2)}$. So the terms
$\exp(m_2 N)$ and $\exp(m_3 N)$ in the solution (\ref{EV}) decay
oscillatingly.
Thus for positive potentials, the scaling solution is stable
whenever this solution exists, whereas
for negative potentials the scaling solution is never stable.

\begin{table}

\begin{tabular}{||c|p{4.02cm}|p{2.8cm}|p{2.5cm}|p{4.2cm}||} \hline \hline
Label & Critical points & Existence & Eigenvalues & Stability \\ \hline
S & $x_1^2+x_2^2=1$, & all $\lambda$ & $(3-\lambda
\sqrt{\frac{3}{2}}x_1)$; &
stable $(\lambda ^2>12)$ \\ 
 & $y_1=y_2=0$ & & $(6-\lambda \sqrt{6}x_2);0$ &
unstable $(\lambda ^2<12)$ \\ \hline
A$_1$, & $(\frac{\lambda}{\sqrt{6}},\sqrt{\pm
(1-\frac{\lambda
^2}{6})},0,0)$, & $\lambda ^2<6$ $(V>0)$ &
$(\lambda ^2-6)/2$; & unstable \\ 
A$_2$ & $(0,0,\frac{\lambda}{\sqrt{6}},\sqrt{\pm
(1-\frac{\lambda ^2}{6})})$ & $\lambda ^2>6$ $(V<0)$ &
$(\lambda ^2-6)/2$; $\lambda ^2$ & \\ \hline
 & $x_1=x_2=\frac{\lambda}{2\sqrt{6}}$, & $\lambda ^2<12 (V>0)$ &
$(\lambda ^2-12)/4$; & stable $(V>0,\lambda ^2<12)$ \\
B & $y_1=y_2=\sqrt{\pm (\frac{1}{2}-\frac{\lambda ^2}{24})}$ &
$\lambda ^2>12 (V<0)$ & $\frac{1}{8}(\lambda ^2-12)\pm$
$\frac{3}{8}\sqrt{(\lambda ^2-4/3)(\lambda ^2 -12)}$ & unstable $(V<0)$
\\ \hline \hline
\end{tabular}

\caption{The properties of the critical points}

\end{table}

The regions of $(\lambda )$ parameter space lead to different
qualitative evolution.
\begin{itemize}
\item For steep positive potentials ($V>0$, $\lambda ^2>12$),
 only a circle $S$ exists, some kinetic-dominated
 scaling solutions of
 which are the late-time attractors. Thus generic solutions in the
 kinetic-dominated regime approach an equal-kinetic-dominated regime,
 the range of which is determined by the value of $\lambda$, at late times.
 That is, the kinetic energy of each field tends to be equal via their
 effect on the expansion.
\item For intermediate positive potentials ($V>0$, $6<\lambda ^2<12$), 
 a circle $S$ and a fixed point $B$ exist. The later is the unique
 late-time attractor. Thus generic solutions start in a kinetic-dominated
 regime and approach the double-kinetic-potential scaling solution.
\item For flat positive potentials ($V>0$, $\lambda ^2<6$),
 all critical points exist. The double-kinetic-potential scaling
 solution is the unique late-time attractor. Thus generic solutions start
 in a kinetic-dominated solution or in a single-kinetic-potential
 solution and approach the double-kinetic-potential scaling solution. 
\item For steep
negative potentials ($V<0$, $\lambda ^2>12$), all critical
 points
exist. The kinetic-dominated scaling solutions are the
 late-time attractors, which correspond to a contracting phase in the
 pre big bang scenario~\cite{PBB}.
\item For intermediate negative potentials ($V<0$, $6<\lambda ^2<12$),
 a circle $S$ and two fixed points $A_1$ and $A_2$ exist.
 There exist no stable points.
\item For flat negative potentials ($V<0$, $\lambda ^2<6$),
 only a circle $S$ exists, which are never stable.
\end{itemize}

We now generalize the above discussion to $m$ scalar fields, by considering
each potential to have a different slope $\lambda _i$
\begin{equation}
V_i(\phi _i)=V_{0i}\exp{(-\lambda_i \kappa \phi _i)}
\end{equation}
Setting $x'_i=0$ and $y'_i=0$, we can get
\begin{eqnarray} 
\label{G5E} 
0&=&-3x_i\left(1-\sum_{j=1}^{m}x_j^2\right)
\pm \lambda_i \sqrt{\frac{3}{2}}y_i^2 \\ 
0&=&y_i\left(3\sum_{j=1}^{m}x_j^2-\lambda_i \sqrt{\frac{3}{2}}x_i\right) 
\label{G6E} 
\end{eqnarray} 
We only consider the $m$-kinetic-potential scaling solution where
$y_i \neq 0$ and $x_i \neq 0$. Using Eq.(\ref{G6E}), we get
\begin{equation}
\label{LC}
\sum_{j=1}^{m}x_j^2=\frac{\lambda_i x_i}{\sqrt{6}}
\end{equation}
Notice that $\lambda_i x_i / \sqrt{6}= c$ is an invariant. So  
\begin{equation}
\label{LCC}
\sum_{j=1}^{m}x_j^2=6c^2\sum_{j=1}^{m}\frac{1}{\lambda_j^2}
\end{equation}
Comparing Eq.(\ref{LCC}) with Eq.(\ref{LC}) gives
$c=(6\sum_{i=1}^m\lambda _i^{-2})^{-1}$. From Eqs.(\ref{G5E}) and
(\ref{G6E}), we obtain the multi-kinetic-potential scaling solution 
\begin{eqnarray} \label{GS1}
x_i &=& \frac{\sqrt{6}c}{\lambda _i} \\
y_i &=& \sqrt{\pm \frac{6c}{\lambda _i^2}(1-c)} 
\end{eqnarray}
Substituting Eq.(\ref{LC}) into Eq.(\ref{SFP}) gives
\begin{equation}
p = \sum_{j=1}^{m} \frac{2}{\lambda_j^2} \label{GS3}
\end{equation}
which is just the result derived in Refs.\cite{LM}. For $m$ scalar fields
with $\lambda_i=\lambda$, Eqs.(\ref{GS1})-(\ref{GS3}) yield $x_i=\lambda
/(m\sqrt{6})$, $y_i=\sqrt{\pm \frac{1}{m}(1-\frac{\lambda ^2}{6m})}$
and  $p=2m/ \lambda^2$, which are consistent with the above
results. The multi-kinetic-potential scaling solution exists for positive
potentials $(\lambda ^2<6m)$, or negative potentials $(\lambda
^2>6m)$. As long as each potential satisfies $\lambda ^2<2m$, this
power-law solution is inflationary. For the case $m=1$, the
dimensionless constant $\lambda$ must be smaller than $\sqrt{2}$
to guarantee power-law inflation~\cite{HW}. However, presently
known theories yield exponential potentials with $\lambda
>\sqrt{2}$. In such cases multiple scalar fields may proceed
inflation. The reason for this behavior is that while each field
experiences the `downhill' force from its own potential, it feels
the friction from all the scalar fields via their contribution to
the expansion~\cite{LM}.

We have presented a phase-space analysis of the evolution for a spatially
flat FRW universe containing $m$ scalar fields with positive or negative
exponential potentials. As an example we study the problems of the fixed
points and their stability in a two-field model. We find that in the 
expanding universe model with sufficiently flat ($\lambda ^2<12$) positive 
potentials the only power-law double-kinetic-potential scaling solution
is the late-time attractor and the well-known inflationary solution with
$a \propto t^p$ where $p=4/\lambda ^2$. A successful inflation can be
driven by multiple scalar fields with a wide range of values for each
potential slope parameter $\lambda$. We also find that the scaling solution
with steep negative potentials is always unstable in the the expanding
universe models. However, sufficiently steep ($\lambda ^2>12$) negative
potentials have kinetic-dominated solutions with $a \propto t^{1/3}$, which
are always the late-time attractors. It can be known that the kinetic
energy of each field tends to be equal via their effect on the expansion.

We emphasize that we have assumed that there is no direct coupling
between
these exponential potentials. It is worth studying further the case their
potentials have different slopes.
It would else be interesting to study the
multi-field dynamics when a perfect fluid is present and for realistic
cross coupling of the kind
$\exp (\lambda_1 \kappa \phi_1+\lambda_2 \kappa \phi_2)$.

\section*{Acknowledgements}
This project was in part supported by NNSFC under Grant
Nos. 10175070 and 10047004 as well as also by NKBRSF G19990754.

\end{document}